\documentclass[aps,prl,twocolumn,superscriptaddress,nofootinbib,showpacs,showkeys,preprintnumbers]{revtex4-1}
\usepackage{graphicx}  
\usepackage{dcolumn}   
\usepackage{bm}        
\usepackage{amsmath, amsthm, amssymb}
\def\lsim{\mathrel{\rlap{
\lower4pt\hbox{\hskip-3pt$\sim$}}
    \raise1pt\hbox{$<$}}}     
\def\gsim{\mathrel{\rlap{
\lower4pt\hbox{\hskip-3pt$\sim$}}
    \raise1pt\hbox{$>$}}}     
\usepackage{color}

\begin{document}

\title{Directed flow in asymmetric nucleus-nucleus collisions \\
and the inverse Landau-Pomeranchuk-Migdal effect}

\author{V. D. Toneev}

\affiliation{Joint Institute for Nuclear Research, Dubna, Russia}

\author{V. Voronyuk}
\affiliation{Joint Institute for Nuclear Research, Dubna, Russia}
\affiliation{Bogolyubov Institute for Theoretical Physics, Kiev,
Ukraine}

\author{E. E. Kolomeitsev}
\affiliation{Faculty of Natural Sciences, Matej Bel University,
Banska Bystrica, Slovakia}

\author{W. Cassing}
\affiliation{Institute for Theoretical Physics, University of
Giessen, Giessen, Germany}

\begin{abstract}
It is proposed to identify a strong electric field - created during
relativistic collisions of asymmetric nuclei - via the observation
of pseudorapidity and transverse momentum distributions of hadrons
with the same mass but opposite charge. The results of detailed
calculations within the Parton-Hadron String Dynamics (PHSD)
approach for the charge-dependent directed flow $v_1$ are presented
for semi-central Cu+Au collision at $\sqrt{s_{NN}}=200$ GeV
incorporating the inverse Landau-Pomeranchuk-Migdal (iLPM) effect,
which accounts for a delay in the electromagnetic interaction with
the charged degree of freedom. Including the iLPM effect we
achieve a reasonable agreement of the PHSD results for the charge
splitting in $v_1(p_T)$ in line with the recent measurements of the
STAR Collaboration for Cu+Au collisions at $\sqrt{s_{NN}}=200$ GeV
while an instant appearance and coupling of electric charges at the
hard collision vertex overestimates the splitting by about a factor
of 10. We predict that the iLPM effect should practically disappear
at energies of  $\sqrt{s_{NN}} \approx$9 GeV, which should lead to a
significantly larger charge splitting of $v_1$ at the future
FAIR/NICA facilities.
 \end{abstract}

\maketitle

The properties of the very initial degrees of freedom in
ultra-relativistic heavy-ion collisions during the passage time of
the impinging nuclei is presently unknown and the ideas vary from a
color-glass-condensate (CGC) \cite{Larry,Gelis} to a gluon dominated
plasma \cite{Horst} or a longitudinal color field that decays to
strongly interacting partons \cite{PHSDrev}. Various suggestions
have been made to distinguish between such scenarios
\cite{ref1,ref2,ref3}, however, a clear discrimination has not been
achieved yet \cite{Pierre}. It was proposed in
Refs.~\cite{VTV14,HHH14} that a strong electric field -- produced
early by the spectator charges -- could help to clarify the problem
by investigating the charge splitting of the directed flow of
particles with equal mass and opposite electric charge as a
function of rapidity and transverse momentum.

Indeed, it has been demonstrated early in Ref.~\cite{SIT9} that the
collective motion of spectator charges in relativistic heavy-ion
collisions can produce extremely strong electromagnetic fields.
Particularly, in peripheral Au-Au collisions at the center-of-mass
energy $\sqrt{s_{NN}}=$200 GeV the magnetic field in the very
initial interaction state can be as high as $|eB_y| \sim 5
{m_\pi^2c^3}/{\hbar}/\hbar e \sim 5 10^{18}$ Gauss, which is the largest
value reachable at terrestrial conditions and even larger than
magnetic fields in magnetars. However, the subsequent analysis of
Au+Au collisions in the energy range up to the top RHIC energies
revealed no visible effect of strong electromagnetic interactions on
global characteristics and, in particular, on sensitive quantities
such as the directed or elliptic flow. The reason for that is not
the very short interaction time of the electromagnetic field with
the charges of the partonic system, as one might expect naively, but
rather a compensation of electric and magnetic forces in symmetric
systems as found in Ref.~\cite{VTC11}. However, it has been argued
that in asymmetric collisions this compensation effect is largely
suppressed due to the different number of protons in the colliding
nuclei \cite{VTV14,HHH14}. Since the strength of the induced
electric field is strongly asymmetric inside the overlap region, one
may expect to observe an asymmetry in the momentum distributions of
produced charged hadrons.  In particular, in Cu+Au collisions the
directed flow, i.e. the first flow harmonic $v_1=<p_x/p_{T}>$ ($p_x$
denoting the momentum projection on the reaction plane while $p_T$
is the transverse momentum), exhibits a dependence on the charge of
partonic or hadronic particles. This has been shown explicitly in
Ref. \cite{VTV14} within microscopic calculations in the framework
of the Parton-Hadron-String Dynamics (PHSD) approach
\cite{CB-PHSD,CB-PHSD2} for Cu+Au collisions at $\sqrt{s_{NN}}=$200
GeV (cf. Fig. 2 in \cite{VTV14}) where one finds a strong electric
field in the central region of the overlap area which is directed
from the Au nucleus to the Cu nucleus.

Detailed calculations of the
directed flow $v_1$ for $\pi^{\pm},\ K^{\pm},\ p$\ and  $\bar p$\ at
the energy $\sqrt{s}=$ 200\,GeV have been carried out in
Ref.~\cite{VTV14} taking into account the influence of the retarded
electromagnetic field created by spectators on the particle
trajectories. The PHSD calculations have been performed also for
Cu+Au collisions for the NICA energies of $\sqrt{s_{NN}}=9$ GeV and
$5$\,GeV. Here the charge-dependent separation effect may be
observed also at 9\,GeV as clearly as at 200\,GeV, however, it
becomes much weaker for $\sqrt{s_{NN}}=5$\,GeV~\cite{TRV16}.

\begin{figure}[thb]
\includegraphics[height=5.50truecm] {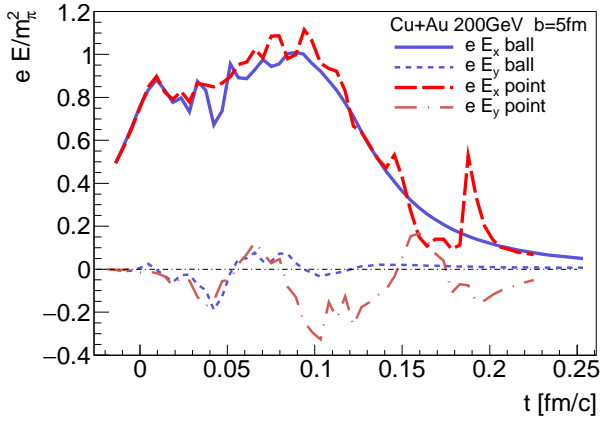}
\caption{ Time dependence of the electric field components $E_x$ and
$E_y$  as generated by point-like charges (dash-dotted line) in the
central point of the overlap region $(x,y,z)=(1,0,0\,{\rm fm})$ for
Cu$+$Au at $b=$5\,fm and $\sqrt{s_{NN}}=$200\,  GeV. The solid and
dotted lines correspond to the fields from the Lorentz contracted
ball-like charge distributions (see text).}
 \label{diagCuAu200-E}
\end{figure}

As noted in Ref.~\cite{VTC11}, the electromagnetic field (EMF) is formed
predominantly by charged spectators at the early stage of the collision during
the passage time of the two colliding nuclei.
Since the number of spectator nucleons decreases with decreasing
impact parameter $b$, the electromagnetic fields should also
decrease gradually with increasing centrality. However, as found in Ref.~\cite{VTC11}
the strength of the average $E_x$ component of the electric field
does not change much in the interval of $b=3-7$ fm.

As seen from the time evolution of the electric field for Cu$+$Au at
$b=$5\,fm and $\sqrt{s_{NN}}=$200\,  GeV in
Fig.~\ref{diagCuAu200-E}, the average strength of the dominant
component  $\langle E_x\rangle$ fields reaches maximal values of
$\langle eE_x\rangle\approx 1.0\, {\rm m^2_{\pi}\, c^3/\hbar {\rm GeV}^2}$ for
a time of $\sim$ 0.15 fm/c which is about the passage time of the
two nuclei. The other components are practically negligible.

\begin{figure}
\centering
\includegraphics[width=0.38\textwidth]{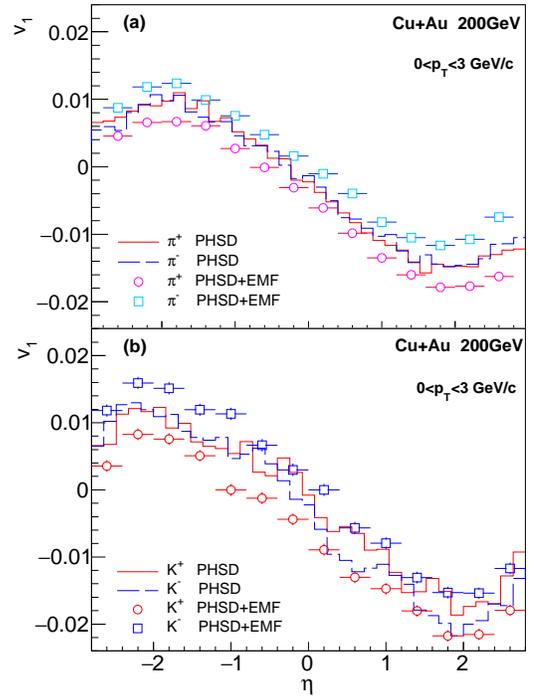}\\
\caption{Pseudorapidity distributions within the pseudorapidity interval
$|\eta|<3$ for positive and negative pions (a) and kaons (b) created in Cu+Au
collisions at $\sqrt{s_{NN}}=$200 GeV in the impact parameter interval
4.4-9.5 fm. The histograms are the result of the standard PHSD transport 
approach without EMF, the symbols are obtained when the electromagnetic 
force is additionally included. }
 \label{v1eta}
\end{figure}

To investigate the influence of the EMF we have calculated within the PHSD approach
various characteristics of the asymmetric Cu+Au collisions at
$\sqrt{s_{NN}}=200$\,GeV. In Fig.~\ref{v1eta} the directed flow $v_1$
is presented as a function of pseudorapidity $\eta$ for charged pions
and kaons. We see that within the pseudorapidity window $|\eta|<3 $
the $\eta$ distributions for $\pi^+(K^+)$ and $\pi^-(K^-)$ are very
close to each other when discarding the EMF in the dynamics. We recall that
the difference increases for larger rapidities and becomes sizable only
for forward or backward rapidities  $|\eta|>3$ which can be attributed  to
a difference in the production mechanism of these mesons~\cite{VTV14}.
The inclusion of the EMF, however, leads to a sizable
separation of these distributions for opposite charges.

\begin{figure}[hb]
\centering
\includegraphics[width=5cm,height=4cm]{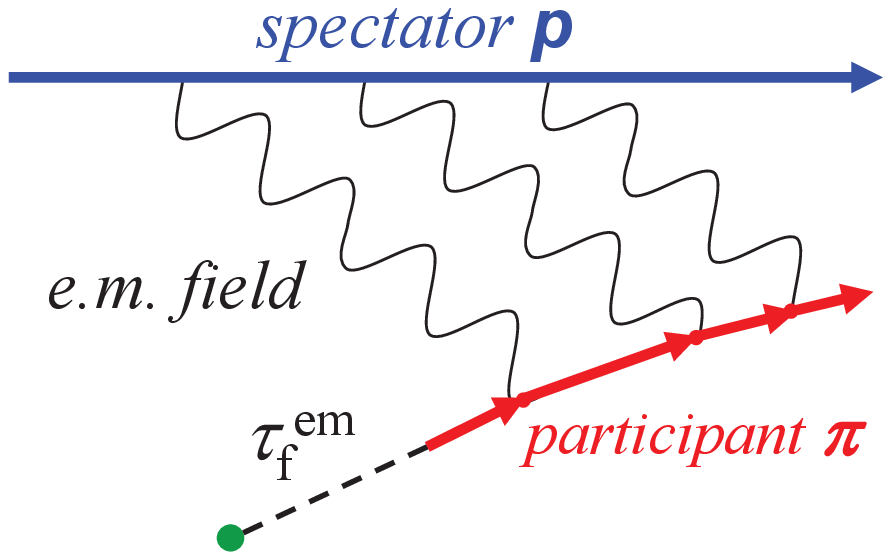}
\caption{The scheme of the inverse LPM effect. The dashed line illustrates the
insensitivity to the electromagnetic field during the formation time for $\tau^{em}_f=(1/10)\tau_f$ of a participant $\pi$, the wavy lines denote the 
electric field.}
 \label{scLPM}
\end{figure}

Although two years have passed since the start of the data-taking for Cu+Au
collisions at
\begin{figure*}
\centering
\includegraphics[width=0.85\textwidth]{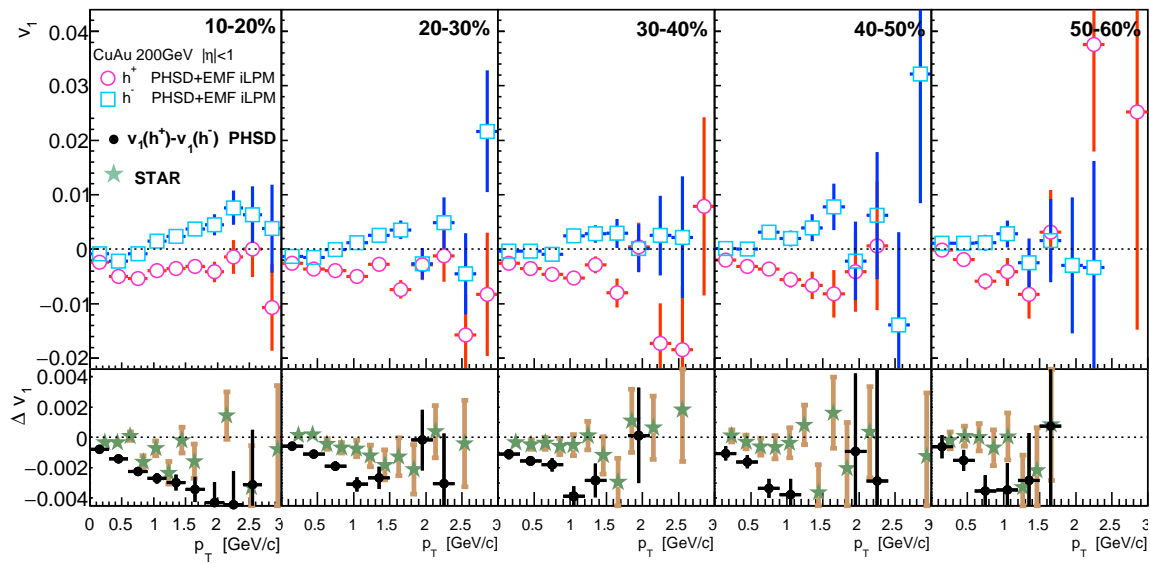}
\caption{Charge-dependent  $p_T$ distributions of positive and negative
hadrons (top) and their difference $\Delta v_1$ (bottom) from asymmetric
Cu+Au collisions at $\sqrt{s_{NN}}=$200 GeV and various centralities including
 the inverse LPM effect. The experimental data (stars) are taken from Ref. \cite{Nii16}.}
\label{v1pt-comp}
\end{figure*}
$\sqrt s_{NN}=200$\,GeV, the first preliminary data on the
charge-dependent anisotropic flow have been reported only 
recently~\cite{Nii16,STAR}. The comparison between the PHSD 
results and the data shows that our calculations overestimate 
the measured splitting in the directed flow of charged 
particles, $$\Delta v_1=v_1(h^+) - v_1(h^-),$$ by a factor of 
about ten. Thus, it is necessary to figure out possible
mechanisms which can reduce $\Delta v_1$ within the  PHSD model:

\noindent i) {One should note first, that the EM field variation
with time could be too fast such the classical treatment of the EMF
is not allowed. For example, according to~\cite{LLIV} the amplitude
of the electric field $eE$ should by larger than the critical field
$e\,E_{\rm crit}=e\sqrt{\hbar c}/(c\Delta t)^2$, where $\Delta t$ is a 
typical time of the field variation.  Since EMFs are described by the
Li\'enard-Wiechert potentials we can estimate the field variation
time for  ultra-relativistic collisions as $\Delta t=E/\dot{E}\sim
\langle{b}\rangle/c$, where $\langle{b}\rangle$ the average impact
parameter of the collision, then we get $e\,E_{\rm
crit}=0.17m_\pi^2/c^3\hbar/(b/{\rm fm})^2$. Therefore, for the typical impact
parameter range considered, the field strength $\gsim 1m_\pi^2$ as
shown in Fig.~\ref{diagCuAu200-E} is large enough to treat the EMF
classically.}\\

\noindent ii) {In our treatment we have considered the charged
particles as point-like and, therefore, the Coulomb interaction
becomes singular at the charge-location. Thus we have recalculated
Cu+Au collisions assuming that the spectator charges have the shape
of a Lorentz-contracted ball. The corresponding results are shown in
Fig.~\ref{diagCuAu200-E} and  we find that the event-averaged  field
strengths  do not change much. Hence,  this modification can not
explain the observed discrepancy.}\\

\noindent iii) {The  analysis of ultrarelativistic elastic $pp$
scattering revealed that at $\sqrt{s_{NN}} \gsim70$\ GeV a
transition could occur from a ball to a hollow toroidal-like
shape~\cite{Dr16}. This certainly may influence the created
electromagnetic field but the scale of this effect is about the same
as the  change of the point-like charge by the ball-like charge as
discussed above.}\\

iv) {A large electric conductivity $\sigma$ and large chiral
magnetic conductivity $\sigma_\chi$ might have some impact on the EM
fields. However, as shown in Ref.~\cite{LSW16} this also should have
a small effect on the retarded electric and magnetic fields created
in heavy-ion collisions. Anyhow, the electric conductivity is
expected to be rather low in the strong QGP \cite{Cass13}.}

Some stronger effects on the $v_1$ splitting might be expected from
changes in the interaction of charges with the electric and magnetic
fields. It is well known that the radiation of photons by high
energy electrons passing through  matter is suppressed for  photons
with a wave lengths larger than the electron mean-free path. For
such wave lengths a transition occurs from an incoherent radiation
of photons in each electron interaction in matter to a coherent
radiation from many interactions. This is the
Landau-Pomeranchuk-Migdal (LPM) effect predicted first in
Ref.~\cite{LP53} and described in a fully quantum-mechanical manner
in Ref.~\cite{Mi56}. In terms of non-equilibrium Green's function
the LPM effect has been reconsidered in Refs~\cite{KV95,KV96}. This
effect can be interpreted as a time delay for an electron after a
collision  before it can fully participate in the electromagnetic
interactions again. In applications to hadron physics the same
arguments were used first by Pomeranchuk and Feinberg in
Refs.~\cite{PF53,FP56}. Later, Feinberg in Ref.~\cite{F66} argued
that after a hard interaction a charged particle "shakes off" its
field and stays in a state, in which its subsequent interactions
differ from the normal one for some time delay until the field is
reestablished. We note that the suppression of soft photon
production in relativistic heavy-ion collisions also has been
analyzed in Ref. \cite{Linnyk} and the LPM effect has been
parameterized in terms of the inverse interaction rate.

\begin{figure*}
\centering
\includegraphics[width=0.44\textwidth]{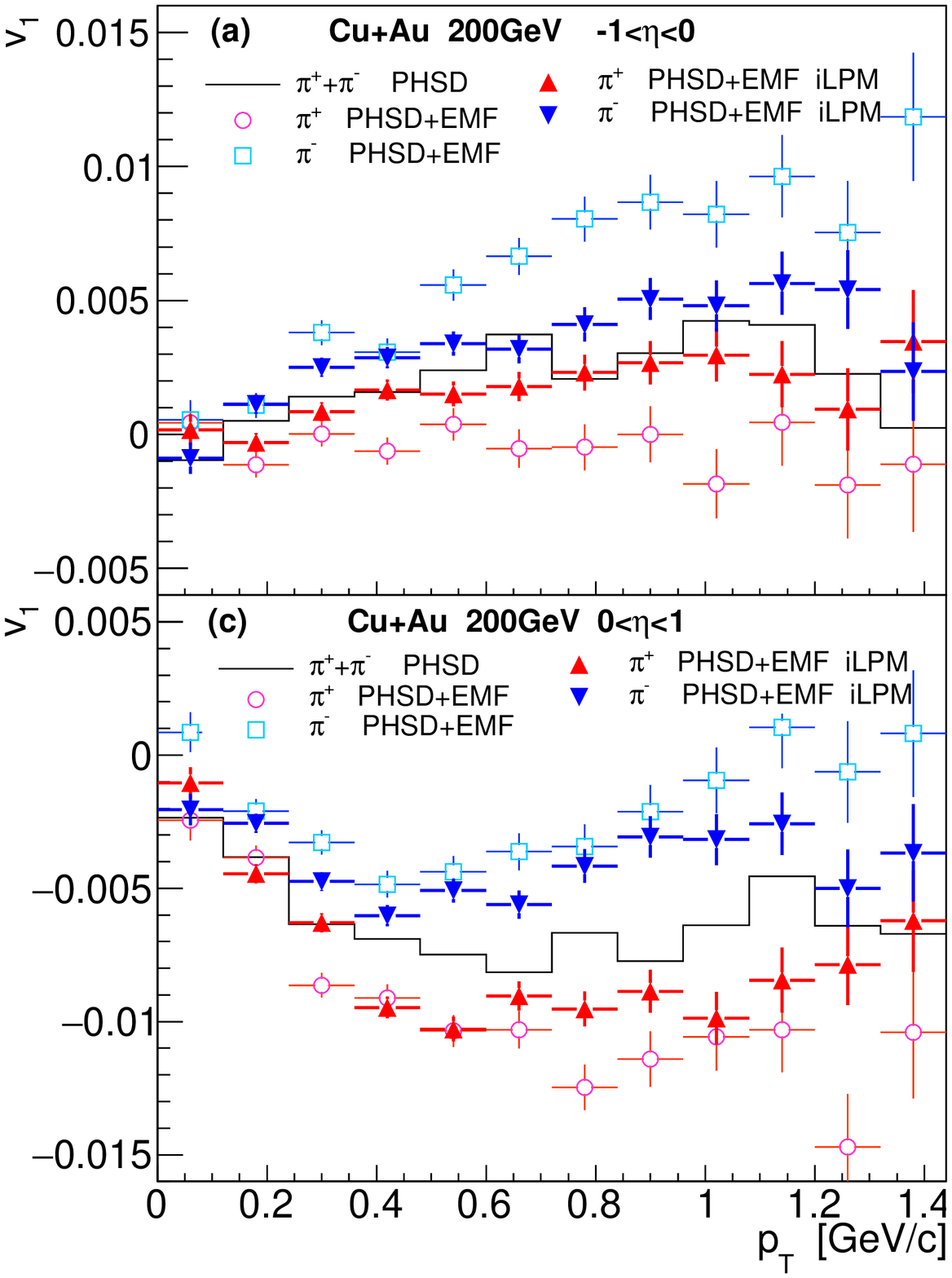}\hspace{7mm}
\includegraphics[width=0.44\textwidth]{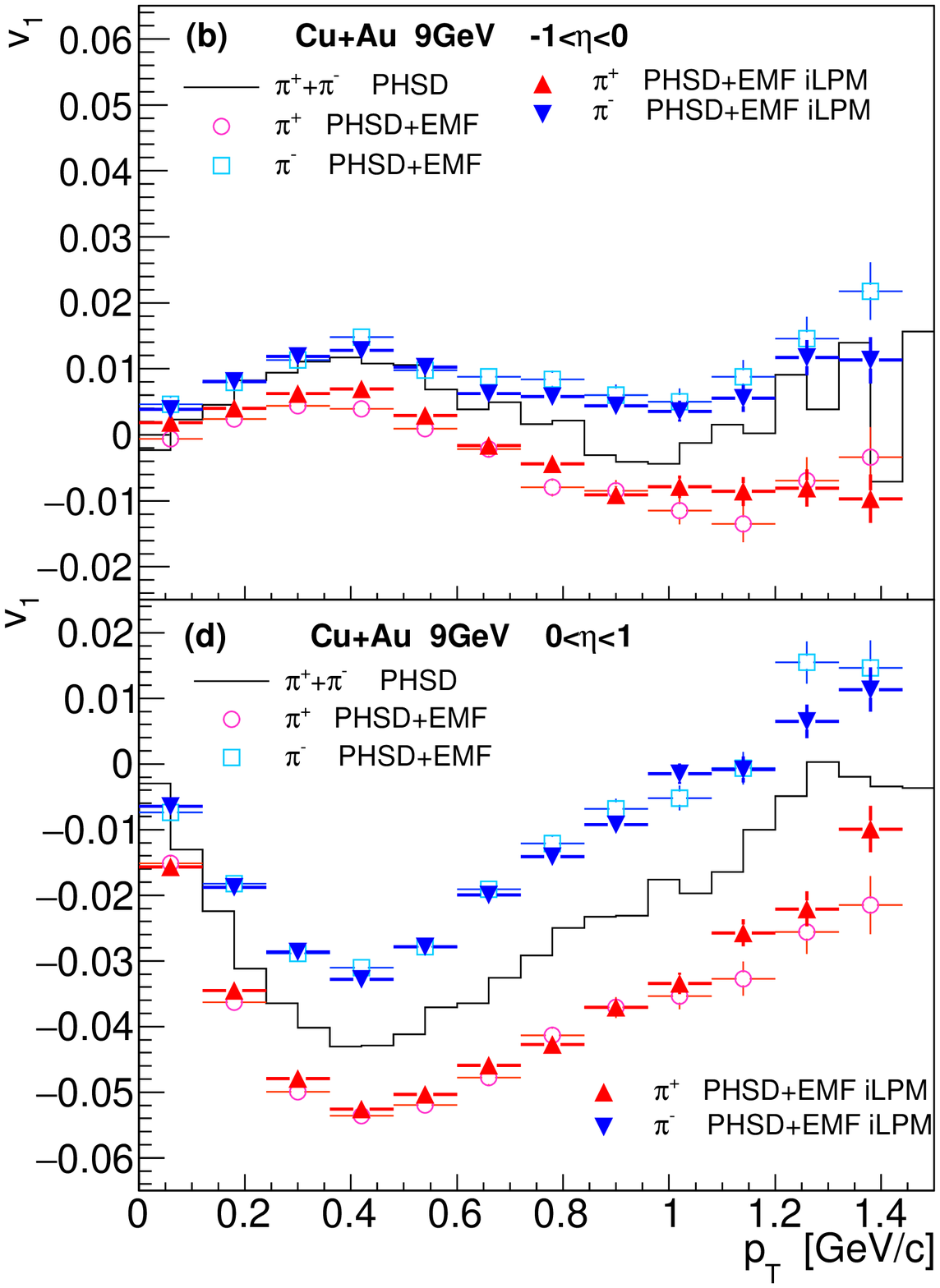}
\caption{Charge-dependent transverse momentum $p_T$ dependence of
the directed flow for pions from asymmetric Cu+Au collisions at
$\sqrt{s_{NN}}=$ 200 GeV (left) and $\sqrt{s_{NN}}=$9 GeV (right).
The backward  and forward emitted pions are plotted in panels (a)
and (c) for 200 GeV and (b) and (d) for 9 GeV, respectively. 
The inverse LPM effect is taken into account with 
$\tau_f^{\rm em}=(1/10)\tau_f$ and shown by full symbols. Other 
parameters employed and the notation are as in Fig.~\ref{v1eta}.}
 \label{v1pt200}
\end{figure*}

A similar concept is inherent in the Lund string model~\cite{BG87}
which incorporates a simple anzatz for the formation time
$\tau_f={\hbar} c E_h/m_T^2$ for quark-antiquark pairs with transverse
mass $m_T$ and energy $E_h$  as well as for the formation of new hadrons  while
disregarding a formation time for leading particles (cf. the
review~\cite{NNN81} where the formation-time concept for hadrons and
the physics of the LPM effect are considered on the same ground in
their different applications). We recall that this formation concept
is also employed in the PHSD approach with a hadronic formation time
$\tau_0\approx 0.8\,{\rm fm}/c$ (in the hadron rest frame), which
allows for a good description of the hadron multiplicities in
heavy-ion collisions in the large energy range from $\sqrt{s_{NN}}=$
3 GeV to 5 TeV \cite{PHSDrev}.

In Ref.~\cite{VTC11} the PHSD model was generalized to take into
account the coupling of a moving charged particle with the generated
electric and magnetic fields. The formation time concept was taken
into account in the particle dynamics such that the generation of
electromagnetic fields only occurs  from formed particles,
dominantly spectator protons. Then, this radiation is traced in
space-time towards a point where it meets a participant charged
particle. This particle may be formed or not yet (the latter case is
shown by the dashed line in Fig.~\ref{scLPM}).  A priori, it is not
evident how the particle will respond to the field under this
conditions. In our early calculations~\cite{VTV14} we assumed that
the EMF acts in the same way on both formed and preformed charged
particles, i.e. $\tau_f^{\rm em}=0$. This assumption is illustrated
in Fig.~\ref{v1eta}. As noted above, these calculations strongly
overestimate the charge splitting of $v_1$ compared to the measured
data.

In the opposite limiting case, when there is no influence of the EMF
on a preformed propagating electric charge (shown by histograms in
Fig.~\ref{v1eta}), no $v_1$ splitting is seen for particles with
opposite electric  charges. (This result was obtained on the statistical
level of about $10^6$ events). Note that the influence of the electromagnetic
field on the conserved charge of a particle in the preformed state looks
like \emph{the inverse LPM effect}.

An intermediate case is presented in Fig.~\ref{v1pt-comp}. Here it
is assumed that the electric field starts to act on the preformed
electric charge with a delay of  $\tau_f^{\rm em}=\tau_f/10$. As
seen, in this case the charge splitting $\Delta v_1$ is in a
reasonable agreement with experimental data\footnote{In different
publications the directions of the bombarding Au or Cu nuclei are
inverted.}. No free normalization factor is used here which implies
that "preformed" charged particles "see" the electromagnetic field
long before being completely formed, i.e. for times $ t<\tau_f /10\,$.

The transverse momentum ($p_T$) dependencies of the directed flow $v_1$
of pions (created in Cu+Au at $\sqrt{s_{NN}} =$200 GeV)  are shown in
Fig.~\ref{v1pt200}. The shape of the $p_T$ spectra  in the forward ($\eta>0$) (b)
and backward ($\eta<0$)(a) directions  are noticeably different.
Without the EMF effect the $v_1(p_T)$ dependence
varies between 0.5\%--1\% in the absolute magnitude
(solid lines in Fig.~\ref{v1pt200}). The inclusion of the EMF splits the
distributions pushing the $v_1(\pi^+)$ upward and  $v_1(\pi^-)$ downward
with respect to the case without  EMF. The charge splitting 
$\Delta v_1$ becomes larger with  increasing transverse momentum
$p_T$. We note that an additional implementation of the iLPM effect
at the top RHIC energy strongly suppresses  the directed flow in
the backward direction but only moderately influences the forward 
component.

We now consider the NICA energy range where the particle creation
occurs at a high baryon density or a large baryonic chemical
potential $\mu_B$. The maximal average energy density reached in a
central  cylinder with radius $R=2$ fm and length $|z|<2.5/\gamma$
fm (where $\gamma \approx \sqrt{s_{NN}}/2m_N$ is the Lorentz factor of
colliding nuclei) is about 1.6 ${\rm GeV/fm^3}$ for a collision at
$\sqrt{s_{NN}}$= 9\,GeV, which implies that a sizeable volume gets
converted to partonic degrees-of-freedom during the collision. In
addition to $\mu_B$, the electric charge chemical potential $\mu_e$
is also important since we are interested in hadrons with opposite
electric charges.

The transverse momentum ($p_T$) dependence of the directed flow
$v_1$ of pions (created in Cu+Au at $\sqrt{s_{NN}} =$9 GeV)  is
shown in Fig.~\ref{v1pt200} for forward (d) and backward (b)
rapidities. As in case of collisions at the top RHIC energy the
created EM field produces  an essential charge splitting of
$v_1(p_T)$, however, an extra inclusion of the iLPM practically does
not effect the $v_1(p_T)$ distributions due to a much longer passage
time of the nuclei and the low delay time $\tau_f^{\rm em}$ due to
the iLPM effect. Note also that the magnitude of the directed flow
at 9 GeV is much higher than that at 200 GeV which opens up
interesting perspectives for lower beam energies (at FAIR/NICA).

From the present study we conclude that the experimental observation
of a charge-dependent splitting of pseudo-rapidity and transverse
momentum distributions in the directed flow provides experimental
evidence for the early creation of strong electromagnetic fields in
relativistic heavy-ion collisions. When accounting for the inverse
LPM (iLPM) effect the coupling of unformed charged partons becomes
slightly delayed and suppresses the charge splitting $\Delta v_1$ in
asymmetric nuclear collisions. The decoherence time ($\tau_f^{\rm
em} \sim \tau_f/10$) allows  to reconcile the PHSD results with the
preliminary experimental observations at the top RHIC energy by the
STAR Collaboration. We predict that the inverse LPM effect should
practically disappear  at energies of  $\sqrt{s_{NN}} \approx$9 GeV,
which leads to a significantly larger charge splitting of $v_1$ at
energies in the BESII program at RHIC
and at the future FAIR and NICA facilities.\\[5mm]

{\bf Acknowledgement}\\
The authors are thankful to D. Voskresensky for useful discussions. The work by E.E.K
was supported in part by the Slovak Grants No. VEGA-1/0469/15. Also, E.E.K.
thanks the Laboratory of Theoretical  Physics at JINR (Dubna) for the warm
hospitality and acknowledges the support by a grant of the Plenipotentiary of
the Slovak Government to JINR.


\begin{thebibliography}{99}

\bibitem{Larry}
E. Iancu, A.  Leonidov, and  L. D. McLerran,
 Nucl. Phys. A {\bf 692}, 583 (2001).
\bibitem{Gelis}
F. Gelis, E. Iancu, J. J.-M. Marian, and R. Venugopalan,
 Ann. Rev. Nucl. Part. Sci. {\bf 60}, 463 (2010).
\bibitem{Horst}
H. St\"ocker  {\it et al.}, J. Phys. G {\bf 43}, 015105 (2016).
%
\bibitem{PHSDrev} O. Linnyk, E. Bratkovskaya, and W. Cassing, Prog. Part. Nucl. Phys {\bf 87}, 50 (2016).

\bibitem{ref1}
V. Vovchenko {\it et al.},  Phys. Rev. C {\bf 94}, 024906 (2016).


\bibitem{ref2}
P. Bozek, A. Bzdak, and V. Skokov,  Phys. Lett. B {\bf 728}, 662
(2014).

\bibitem{ref3}
V.P. Konchakovski, W. Cassing, and V.D. Toneev,  J. Phys. G {\bf
41}, 105004 (2014).

\bibitem{Pierre} P. Moreau,  O. Linnyk, { W. Cassing}, and E. L.
Bratkovskaya,
Phys. Rev. C {\bf 93}, 044916 (2016).
%
\bibitem{VTV14} V. Voronyuk, V. D. Toneev, S. A. Voloshin and W. Cassing,
Phys. Rev. C {\bf 90}, 064903 (2014).
%
\bibitem{HHH14}  Y. Hirono, M. Hongo and T. Hirano, Phys. Rev. C {\bf 90}, 021903(R) (2014).
%
\bibitem{SIT9}
V. Skokov, A. Illarionov and V. Toneev, Int. J. Mod. Phys. A {\bf 24}, 5925 (2009).
%
\bibitem{VTC11} V. Voronyuk {\it et al.},
 Phys. Rev. C {\bf 83}, 054911 (2011).
%
\bibitem{CB-PHSD}
W. Cassing and E. Bratkovskaya, Phys. Rev. C {\bf 78}, 034919
(2008); Nucl. Phys. A {\bf 831}, 215 (2009).
%
\bibitem{CB-PHSD2}
E. Bratkovskaya {\it et al.}, Nucl. Phys. A {\bf 856}, 162 (2011).
%
\bibitem{TRV16} V. Toneev, O. Rogachevsky and V. Voronyuk,
Eur. Phys. J., A  {\bf 52}, 264 (2016).

\bibitem{Nii16} T. Niida for the STAR Collaboration,
    arXiv:1601.01017.
\bibitem{STAR}L. Adamchuk {\it at al.} (STAR Collaboration),
    arXiv:1608.04100.
%
\bibitem{LLIV} V.B.~Berestetskii, E.M~Lifshitz, and L.P.~Pitaevskii, Quantum Electrodynamics, Pergamon Press, Oxford (1982), section 5.
%
\bibitem{Dr16} I.M. Dremin, arXiv:1605.08216.
%
\bibitem{LSW16} H. Li, X.-l. Sheng and Q. Wang, 
    arXiv:1602.02223.
%
\bibitem{Cass13} W. Cassing,  O. Linnyk, T. Steinert, and V. Ozvenchuk,
Phys. Rev. Lett. {\bf 110}, 182301 (2013).
%
\bibitem{LP53} L.D. Landau and I.Ya. Pomeranchuk, Dokl. Akad. Nauk SSSR {\bf 92}, 535 (1953); {\bf 92}, 735 (1953).
%
\bibitem{Mi56} A.B. Migdal, Phys. Rev. {\bf 103}, 1811 (1956).
%
\bibitem{KV95} J.~Knoll and D.N.~Voskresensky, Phys. Lett. B {\bf 351}, 43 (1995).
%
\bibitem{KV96} J.~Knoll and D.N.~Voskresensky, Ann. Phys. {\bf 249}, 532 (1996).
%
\bibitem{PF53} I.Ya. Pomeranchuk and E.L. Feinberg , Dokl. Akad. Nauk SSR {\bf 93}, 439 (1953).
\bibitem{FP56} E.L. Feinberg and I.Ya. Pomeranchuk, Nuovo Cim. Suppl. {\bf 3}, Ser. 10, 652  (1956).
%
\bibitem{F66} E.L. Feinberg, Sov. Phys. JETP {\bf 23}, 132 (1966);
%
\bibitem{BG87} A. Bialas and M. Gyulassy, 
    Nucl. Phys. B {\bf 291}, 793 (1987).
%
\bibitem{NNN81} N. N. Nikolaev, 
Usp. Fiz. Nauk {\bf 134}, 369 (1981); translated as
Sov. Phys. Usp. {\bf 24}, 531 (1981).
%
\bibitem{Linnyk}   O. Linnyk {\it et al.},
Phys. Rev. C {\bf 92}, 054914 (2015).







\end{thebibliography}
\end{document}